\documentclass[%
 aip,
 amsmath,amssymb,
 reprint,%
]{revtex4-1}

\usepackage{graphicx}
\usepackage{dcolumn}
\usepackage{bm}

\usepackage[utf8]{inputenc}
\usepackage[T1]{fontenc}
\usepackage{mathptmx}
\usepackage{xcolor}
\usepackage{url}            
\usepackage{hyperref}       

\DeclareMathOperator{\arctanh}{arctanh}
\DeclareMathOperator{\sign}{sign}
\DeclareMathOperator{\Tr}{Tr}
\newcommand{\correct}{\textcolor{black} } 

\begin{document}

\preprint{AIP/123-QED}

\title[1D Spin-Crossover Molecular Chain with Degenerate States]{1D Spin-Crossover Molecular Chain with Degenerate States}

\author{Andrii \surname{Gudyma}}
\affiliation{Max Planck Institute of Microstructure Physics, Weinberg 2, 06120 Halle/Saale, Germany}

\author{Iurii \surname{Gudyma}}%
 \email{yugudyma@gmail.com}
\affiliation{Physical, Technical and Computer Sciences Institute of Yuriy Fedkovych Chernivtsi National University, 58012 Chernivtsi, Ukraine}%

\date{\today}

\begin{abstract}
A study of the one-dimensional molecular chain (MC) with two single-particle degenerate states is presented.
We establish connection of the MC with the Ising model with phononic interactions and investigate properties of the model using a transfer matrix method. 
The transfer matrix method offers a promising pathway for simulating such materials properties.
The role of \correct{degeneracy} of states and phononic interaction being made explicit. 
We analyze regimes of the system and parameters of the occurring crossover.  
Here, we present exact results for the magnetization per spin, the correlation function and the effective volume of the system.
\correct{We demonstrate possibility of existence of two peaks in the specific heat capacity thermal behavior.}
\end{abstract}

\maketitle

\section{Introduction}
\label{sec:Introduction}

Quasi-one-dimensional systems play an essential role in nanotechnology.
Usually phase transitions do not take place in these systems.
It is well known that, in general, one dimensional systems with short range interaction do not undergo a phase transition, with possible exception when the studied model is considered at zero or infinite temperatures.
The absence of phase transition in such systems has been determined by wide variety of conditions.
It means that the manifestation of each case of phase transition in one-dimensional (1D) systems demands a detailed study.
\correct{For example, presence of hysteresis and many-step crossover~\cite{Spin-Crossover1d-1} were found in SCO materials. 
These phenomena appear as a result of the spin-elastic frustration. } 
Depending on the particular configuration of the magnetic molecules, a variety of magnetization behaviors can be described even with a relatively simple model.
The importance of such studies is related to the occurrence of first-order phase transition in molecular chains, that could find a wide variety of practical applications, including further miniaturization of nanodevices and nanomaterials~\cite{Nanotechnology}.
Molecular building blocks are among the most promising candidates for the future electronics~\cite{SENTHILKUMAR2017176}.

When the energy gap between the ground state and the first excited state becomes sufficiently perceptible and \correct{distances to the rest of the excited states are much greater, then the two-state model may be used as the theoretical framework}. 
The two-state interaction model is formulated in terms of Ising spin variables, $s_{n}=\pm 1$. 
Mathematically, this corresponds to the application of the quasi-spin-$\frac{1}{2}$ Ising model for molecular chains.
If the lowest states of the system are degenerate, then we have a degenerate Ising model.
Note that, as a rule, the degeneracy of the high-energy spin state is more significant.
\correct{
Ising model with the degenerated states was introduced by Huang~\cite{Huang1975}, and developed and studied numerically~\cite{HARRIS1985299, bousseksou1992ising, boukheddaden2000one, Schebarchov2014, sienkiewicz2014finite, cajahuaringa2019nonequilibrium}.
The equivalence of the states' degeneracy and temperature dependent effective magnetic field acting on spin-crossover molecules was established~\cite{bousseksou1992ising, boukheddaden2000one}. 
Various mechano-elastic~\cite{nasser2005diluted, konishi2008monte, Lecomte2008Two-variable, enachescu2009model, ye2015monte} and Ising-like models~\cite{linares1999analytical, Boukheddaden_2007, boukheddaden2008molecular} were developed to describe effects which appear in spin crossover materials.
Model which deserve special focus of interest is the compressible Ising model~\cite{zagrebnov1972spin, Salinas_1973, Henriques_1987, lehmann2012anomalous, Spin-Crossover1d-1}.
For this model in 1D the elastic interactions were taken into account using transfer matrix technique~\cite{Linares2004, Boukheddaden2007} and this model was applied to spin crossover materials~\cite{Spin-Crossover1d-1, Spin-Crossover1d-2, Spin-Crossover1d-3}. }

For example, in coordination Fe (II) linear chain compounds exhibiting thermal spin crossover (SCO) transitions are associated with diamagnetic-to-paramagnetic switching between low-spin (LS) and high-spin (HS) state~\cite{nebbali2018one, lan2019thermal, vinogradova2020cooperative}.
In other words, SCO chains undergo drastic changes in the spin configuration, leading them to switch in a reversible way from a low-spin state (LS) to a high-spin state (HS).
The SCO phenomenon represents the paradigm of bistability at molecular level which emerges at the macroscopic scale, offering potential applications in the development of new generations of electronic devices such as molecular spintronics devices, nonvolatile memories, molecular sensors, displays, and reversible switches~\cite{raman2013interface, matsumoto2014programmable, bairagi2016molecular, mullaney2017spin, kipgen2018evolution, ridier2020unprecedented, kobke2020reversible}.
The energy difference between HS and LS states is due to the competition between the crystal field splitting, which prefers doubly occupied $d$ orbitals and, hence, LS, and Hund's first rule, which favors the HS state.
At lower temperature the LS state is dominant, while the HS state is preferred at higher temperature~\cite{halcrow2013spin, gudyma2015kinetics}.
From this point of view, the study of chain Fe~(II) complex by x-ray-absorption fine-structure spectroscopy (XAFS) afford opportunity to obtain detailed information on the phase transition from the local-structure aspect~\cite{yokoyama1998spin}.
The extended XAFS analysis showed the abrupt change of the interatomic distance at the transition temperature.
Large distance differences between the LS and HS states are ascribed to the electron occupation of the $3d$ levels in the HS state.
The LS state shows the $^{1}A_{1g}$ electronic configuration of $(3d t_{2g})^6 (3de_{g})^0$ where no electrons occupy the $\sigma$-antibonding $3de_{g}$ orbitals, while the HS state exhibits the $^{5}T_{2g}$ configuration of $(3d t_{2g})^6 (3de_{g})^2$, this leading to a significant weakening of the metal-ligand bond.
Videlicet, the coupling of the molecular electronic state with lattice transformation together with intermolecular interactions result in very cooperative transitions.

In this work we consider molecular chain of particles with two inner states.
In our model these two inner states may be degenerated.
The particles are a subject to a pair potential of general type which differentiates the inner states of particles.
We establish the connection between molecular chain and the Ising-like model with phonon interactions.
We treat the model analytically using transfer matrix formalism.
Here our goal is first to show the role of degeneracy of states for scenario of spin transition and analyze parameters of the spin crossover.
We focus our attention on the thermal behavior of the magnetization and specific heat capacity.

The outline of this work is as follows.
Sec.~\ref{sec:Hamiltonian} defines the model's formalism.
In Sec.\ref{sec:Transfer_matrix_formalism} we introduce the transfer matrix formalism and make finite $N$ analytic calculations of the effective volume, average magnetization and the correlation function.
In Sec.~\ref{sec:Spin-crossover} we present our analytical and numerical results for spin-crossover molecular chain.
Finally, results and discussions are given in Sec.~\ref{sec:SUMMARY}.

\section{The system}
\label{sec:Hamiltonian}

During the past two decades molecular magnetism has experienced significant advancements, and molecular materials have been able not only to display the different type of magnetic behaviors, initially identified in simpler solids including ferromagnets, but they have also been able to provide examples of materials exhibiting novel magnetic phenomena~\cite{ molnar2018spin, coronado2020molecular}.
In order to characterize the behavior of 1D molecular chain considering both, structural and magnetic properties, we are basing on the simple conception in which the molecules compound a coordination polymer.

\begin{figure*}[!htb]
    \centering
    \includegraphics[width=0.43\columnwidth]{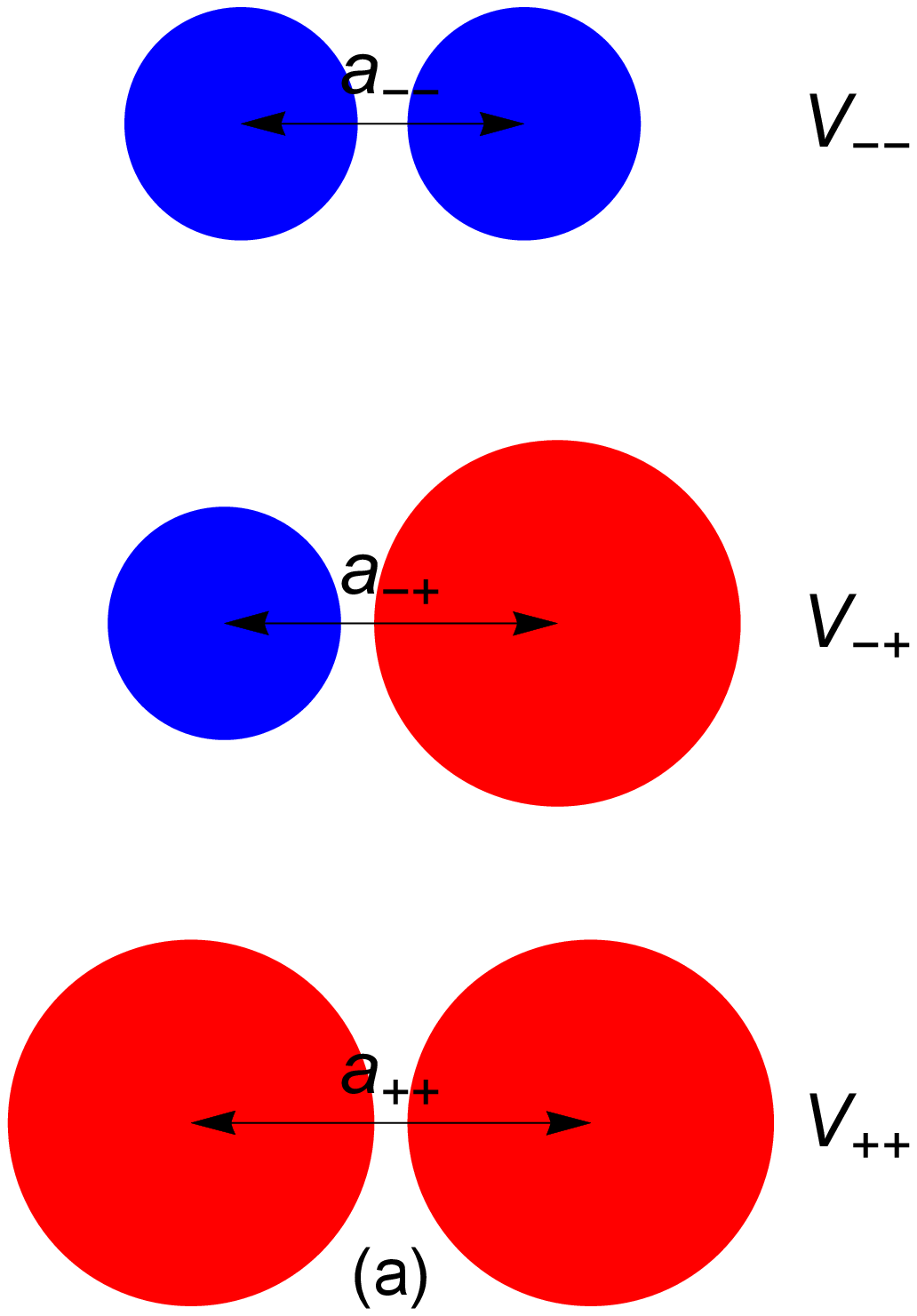}
    \includegraphics[width=1.0\columnwidth]{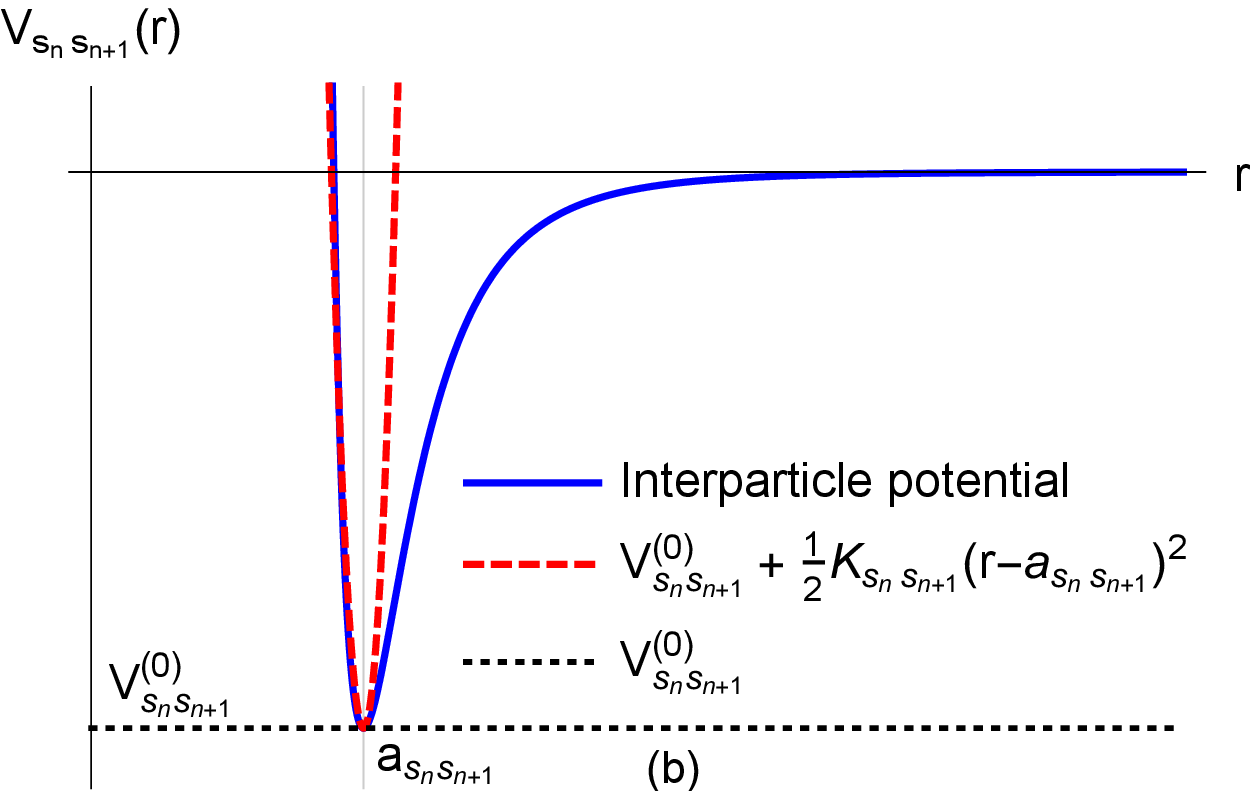}
    \caption{Schematic interactions of the pseudo-spin states and treatment of the inter-particle potential. 
    (a) All possible configurations of the nearest pseudo-spin states. Interaction potentials and average distances between particles depend on the pseudo-spin states. 
    (b) Interaction potential and harmonic approximation. We consider possible displacement of the particles from the equilibrium position for the given pseudo-state configuration to be small.}
    \label{fig:pseudospins_interactions}
\end{figure*}

We denote higher energy state as pseudo-spin $+1$ state and lower energy state as pseudo-spin $-1$ state. 
Spin $+1$ state has degeneracy $g_+$ and spin $-1$ has degeneracy $g_{-}$.
We assume pair interactions depend on the particles pseudo-states, and the potentials of the corresponding pairs of pseudo-spin states interactions are $V_{--}(r)$, $V_{+-}(r)$ and $V_{++}(r)$.
Schematic view of the three microscopic configurations of two neighboring atoms in the chain is shown in the Fig.~\ref{fig:pseudospins_interactions}(a).
The Hamiltonian of the system is a sum of the pair potentials and single particle field\correct{~\cite{Linares2004, Boukheddaden2007, Lecomte2008Two-variable}}
\begin{equation}
    \hat{H} = \sum_{n=1}^{N-1} V_{s_ns_{n+1}}(x_n-x_{n+1}) + \sum_{n=1}^{N} W_{s_n},
    \label{eq:Initial_Hamiltonian}
\end{equation}
where $N$ is the total number of molecules and $W_{s_n}$ is the energy of the single-molecule pseudo-state.
The difference of the pseudo-state energies $\Delta = W_{+} - W_{-}$ is the external ligand field acting on a single molecule.
We apply an harmonic approximation for the pair nearest-neighbor potential $V_{s_ns_{n+1}}(r)$ at the potential minimum %
\begin{equation}
    V_{s_ns_{n+1}}(r) = V_{s_ns_{n+1}}^{(0)} + \frac{1}{2} K_{s_ns_{n+1}} \left(r-a_{s_ns_{n+1}}\right)^2,
    \label{eq:V_ss_n}
\end{equation}
where $a_{s_ns_{n+1}}$ is a distance where potential has minimum, $V_{s_ns_{n+1}}^{(0)}$ is the potential depth and $K_{s_ns_{n+1}}$ is an elastic constant coupling $n$-th and $(n+1)$-st molecules in the pseudo-states $s_n$ and $s_{n+1}$.
Schematic treatment of the potentials is shown in Fig.~\ref{fig:pseudospins_interactions}(b).
In Eq.~(\ref{eq:V_ss_n}) the first term cannot be nullified as $V_{s_ns_{n+1}}^{(0)} = V_{s_ns_{n+1}}(a_{s_ns_{n+1}})$ and it is different for the $V_{--}(r)$, $V_{+-}(r)$ and $V_{++}(r)$ potentials.

Let's introduce variables $q_n = x_n-x_{n+1}$.
We split initial Hamiltonian~(\ref{eq:Initial_Hamiltonian}) to two parts 
\begin{equation}
    \hat{H} = \hat{H}_1 + \hat{H}_2,
    \label{eq:Separated_Hamiltonian}
\end{equation}
where
\begin{equation}
    \hat{H}_1 = \sum_{n=1}^{N-1} V_{s_ns_{n+1}}^{(0)}  + \sum_{n=1}^{N} W_{s_n},
    \label{eq:Separated_Hamiltonian_1}
\end{equation}
and
\begin{equation}
    \hat{H}_2 = \frac{1}{2} \sum_{n=1}^{N-1} K_{s_ns_{n+1}} \left(q_n-a_{s_ns_{n+1}}\right)^2 .
\end{equation}

We rewrite first part of the Hamiltonian (\ref{eq:Separated_Hamiltonian_1}) in terms of pseudo-spins
\begin{equation}
    \hat{H}_1 
    =  E_0\! -\! \sum_{n=1}^{N-1}\! J\! s_n s_{n+1}\! -\! \sum_{n=1}^{N-1}\! B\! \frac{s_n+s_{n+1}}{2}\! -\! W(s_1)\! -\! W(s_N)\!,
\end{equation}
where 
\begin{subequations}
\begin{equation}
    E_0 = \frac{N-1}{4} \left( V_{--}^{(0)} + V_{++}^{(0)} \right) + \frac{N-1}{2} V_{+-}^{(0)} + N \frac{W_{+} + W_{-}}{2},
\end{equation}
\begin{equation}
    J = - \frac{1}{4} \left( V_{--}^{(0)} + V_{++}^{(0)} \right) + \frac{1}{2} V_{+-}^{(0)},
\end{equation}
\begin{equation}
    B = \frac{1}{4} \left( V_{++}^{(0)} - V_{--}^{(0)} \right) - \frac{\Delta}{2},
\end{equation}
\begin{equation}
    W(s_n) = - \frac{\Delta}{4} s_n.
\end{equation}
\label{eq:Ising_model_parameters}
\end{subequations}
The spins on the edge of the system experience additional external field $W(s)$ due to the geometry of the system.

\subsection{Partition function}
\label{sec:partition_function}

The statistical properties of the model are completely determined by the partition function
\begin{multline}
    Z\! = \! \sum_{\langle s_1, \ldots, s_N \rangle}\! \iiint dq_1 \cdots dq_{N-1} g_{s_1} \cdots g_{s_N} e^{-\beta E\left( q_1, \ldots, q_{N-1}, s_1, \ldots, s_N \right)}
    \\ = \sum_{\langle s_1, \ldots, s_N \rangle} g_{s_1} \cdots g_{s_N} 
    \left( \prod_{n=1}^{N-1} 
     \sqrt{\frac{2 \pi}{\beta K_{s_ns_{n+1}}}}  \right)  e^{-\beta E_1} ,
    \label{eq:partition_function}
\end{multline}
where the sum is over all states $\langle s_1, \ldots, s_N \rangle$ with energy $E\left( s_1, \ldots, s_N \right)$, $g_{s_n}$ represents the degeneracy of the state $s_n$ and $\beta$ denotes the inverse of the Boltzmann constant times temperature.
It is convenient to consider an ensemble in which $Z$ depends on the temperature and the field.
Certainly, the choice of boundary conditions becomes irrelevant in the thermodynamic limit.

The coefficients we get during the integration over the phonon degrees of freedom may be expressed in terms of the spin operators as follows
\begin{equation}
    \sqrt{\frac{2 \pi}{\beta K_{s_ns_{n+1}}}} = e^{\delta \epsilon + \delta b (s_n+s_{n+1})/2 + \delta j s_n s_{n+1}},
\end{equation}
where
\begin{subequations}
\begin{equation}
    \delta \epsilon = - \frac{1}{8} \ln \left( \frac{\beta^4}{(2 \pi)^4} K_{+-}^2 K_{--} K_{++}  \right),
\end{equation}
\begin{equation}
    \delta j = \frac{1}{8} \ln \left(  \frac{K_{+-} ^2}{ K_{--} K_{++} }  \right),
\end{equation}
\begin{equation}
    \delta b = \frac{1}{4} \ln \left( \frac{K_{--}}{ K_{++} } \right).
\end{equation}
\label{eq:Ising_model_corrections}
\end{subequations}
We present spin state degeneracies as follows
\begin{equation}
    g_{s_n} 
    = e^{ \frac{1}{2} \left( \ln  g_{+} + \ln  g_{-} \right) + \frac{1}{2} \left( \ln  g_{+} - \ln  g_{-} \right) s_n}.
\end{equation}
We express expression under the exponent of the (\ref{eq:partition_function}) as a sum of pair terms and single-particles terms acting only on the boundary.
Therefore we have 
\begin{equation}
    Z = \sum_{\langle s_1, \ldots, s_N \rangle}  e^{\epsilon +\sum_{n=1}^{N-1} v(s_n,s_{n+1}) + w(s_1) + w(s_N)  }
    \label{eq:partition_function_free_boundary}
\end{equation}
where 
\begin{equation}
    w(s_{n}) = w \frac{ s_{n}}{2},
\end{equation}
with $w = \frac{1}{2} \ln g - \frac{ \beta \Delta }{2}$, and effective two-particle energy terms 
\begin{equation}
    v(s_n,s_{n+1}) =  j s_n s_{n+1} + b (s_n + s_{n+1})/2,
\end{equation}
and 
\begin{subequations}
\begin{equation}
    \epsilon  = - \beta E_0 + \frac{N}{2} \left( \ln  g_{+} + \ln  g_{-} \right) + (N-1) \delta \epsilon,
\end{equation}
\begin{equation}
    j  = \beta J + \delta j,
\end{equation}
\begin{equation}
    b  = \beta B + \frac{1}{2} \ln g + \delta b.
\end{equation}
\label{eq:effective_parameters}
\end{subequations}
We make notation $g = \frac{g_{+}}{g_{-}}$.
As we will see later, it is not the degeneracy of each energy state that is essential, but the ratio of HS and LS states degeneracies.

The partition function (\ref{eq:partition_function_free_boundary}) can be expressed as the partition function of the Ising model with the effective Hamiltonian
\begin{multline}
    \hat{H}_{eff} = E_{0,eff} - \sum_{n=1}^{N-1} J_{eff}  \hat{s}_n \hat{s}_{n+1}  - \sum_{n=2}^{N-1} B_{eff}  \hat{s}_n 
    \\ + \frac{B_{boundary}+B_{eff}}{2} (\hat{s}_1 +\hat{s}_N) , 
\end{multline}
where $E_{0,eff} = E_0 - \frac{N k_B T}{2} \ln  g_{+}  g_{-}  - (N-1) \delta \epsilon k_B T$, $J_{eff} = J \!+\! \delta j k_B T$, $B_{\text{eff}} = B + \frac{k_B T}{2} \ln g + \delta b k_B T $ and $B_{boundary} = -\frac{\Delta}{2} +\frac{k_B T}{2} \ln g$. 
The effective Hamiltonian coincides with the Ising model Hamiltonian with the reference energy, effective magnetic field and ferromagnetic interaction constant being functions of temperature.
This dependence roots from the taking into account pseudo-states degeneracy and phononic interactions.
Eqs.~(\ref{eq:Ising_model_parameters}) and (\ref{eq:Ising_model_corrections}) establish correspondence between the parameters of the actual inter-molecular potentials and parameters of the effective Ising model.

The effective \correct{temperature dependent magnetic field~\cite{bousseksou1992ising, boukheddaden2000one}} $B_{\text{eff}} = B + \frac{k_B T}{2} \ln g + \delta b k_B T $ acts on the bulk, while molecules on the edges are subject of the external field $\frac{1}{2}(B_{boundary}+B_{eff})$, reflecting their boundary distinction.   
Effective Ising-like systems with the phononic interactions were introduced in works~\cite{Linares2004, Boukheddaden2007}.
In our paper we continue investigation of this model with lifting down two significant restrictions: a) all phononic interactions arise from the intermolecular interactions have same potential depth; b) average distances of the intermolecular potential should form arithmetic progression.
We note, this results in having different form of the effective ferromagnetic interaction constant comparing to conclusions~\cite{Linares2004, Boukheddaden2007}.
We consider phononic interactions in the scope of corrections to the parameters of an effective Ising model while coefficients $\delta \epsilon$, $\delta j$ and $\delta b$ define corrections to the effective energy reference, ferromagnetic interaction strength and external field.
Accordingly the fictitious spins interact with their nearest neighbor in the Ising way (parametrized by the phononic interactions).
In addition, we have an external field (B) parametrized by the phononic interactions and the degeneracy between the two lower states.
The spin-independent part cannot be discarded since it is a function of temperature.
Altogether, our model contains the following parameters: the degeneracy between the two lower states $g$, the inter-molecular coupling $J$, the external ligand field $\Delta$ and the external field $B$, the elastic constants $K_{s_ns_{n+1}}$, the temperature $T$.

\section{Transfer-matrix formalism}
\label{sec:Transfer_matrix_formalism}

\correct{Thermodynamic properties of the system are completely described with the partition function.
Here we use the transfer matrix formalism~\cite{Linares2004, Boukheddaden2007} to calculate the partition function.  }
We rewrite the partition function (\ref{eq:partition_function_free_boundary}) as
\begin{equation}
    Z = e^\epsilon \Tr \hat{T}^{N-1} \hat{R} ,
    \label{eq:partition_function_non_periodic}
\end{equation}
where transfer matrix is
\begin{equation}
    \hat{T} = e^{v(s_n,s_{n+1})} =
    \begin{pmatrix}
        e^{ j + b } 
        & e^{ -j  } \\
        e^{ -j  } 
        & e^{ j - b } 
    \end{pmatrix},
\end{equation}
and matrix $\hat{R}$ is accounting effects of the field acting on the surface spins
\begin{equation}
    \hat{R} = e^{ w(s_N) + w(s_1) } 
    =
    \begin{pmatrix}
        e^{  w  } 
        & 1 \\
        1
        & e^{ - w  } 
    \end{pmatrix}.
\end{equation}

For calculating $\Tr \hat{T}^{N-1} \hat{R}$ we should make a rotation of the basis to one where the transfer matrix is diagonal 
\begin{equation}
    Z = e^\epsilon \Tr \hat{U} \hat{U}^{-1} \hat{T}^{N-1} \hat{U} \hat{U}^{-1} \hat{R},
    \label{eq:partition_function_non_periodic_1}
\end{equation}
where the rotation matrix
\begin{equation}
    \hat{U} = 
        \begin{pmatrix}
        \cos \phi
        & \sin \phi \\
        - \sin \phi
        & \cos \phi
    \end{pmatrix},
\end{equation}
and the rotation angle $\phi$ is given by the equation
\begin{equation}
    \cot 2 \phi = e^{2 j  } \sinh (b  ).
\end{equation}
The eigenvalues of the transfer matrix are
\begin{equation}
    \lambda_{\pm} = \left( e^{j } \cosh b  \pm \sqrt{e^{2 j } \sinh^2 b  + e^{-2 j }} \right) .
    \label{eq:lambda+}
\end{equation}
Therefore, we obtain the partition function
\begin{equation}
    Z = e^\epsilon \left( c_{+} \lambda_{+}^{N-1} + c_{-} \lambda_{-}^{N-1} \right),
\end{equation}
where the coefficients are
\begin{subequations}
\begin{equation} 
    c_{+} = \cosh w + \frac{e^{-2j} + \sinh b \sinh w }{\sqrt{\sinh^2 b  + e^{-4 j }}}, 
    \label{eq:c+}
\end{equation}
\begin{equation}
    c_{-} = \cosh w - \frac{e^{-2j} + \sinh b \sinh w }{\sqrt{\sinh^2 b  + e^{-4 j }}}.
\end{equation}
\end{subequations}

The free energy density is given by
\begin{equation}
    f = -\frac{1}{\beta N} \ln Z .
\end{equation}
In the thermodynamic limit we have the desired result:
\begin{equation}
    f = - \lim_{N \to \infty} \frac{1}{\beta N} \ln Z  = - \frac{\epsilon}{N \beta} - \frac{1}{\beta} \ln \lambda_+.
\end{equation}
Average fictitious magnetization per spin can be calculated directly by the formula
\begin{equation}
     \left\langle s \right\rangle = \frac{1}{N} \sum_{n=1}^N \left\langle s_n \right\rangle = \frac{1}{N} \sum_{n=1}^N \frac{1}{Z} \mathrm{Tr} \hat{s}_n e^{-\beta \hat{H}} = \frac{1}{N} \frac{\partial \ln Z}{\partial b } .
\end{equation}
The magnetization per spin $m = \left\langle s \right\rangle$ at nonzero $T$ and $B$: 
\begin{equation}
    m = \frac{\sinh(b )}{ \sqrt{ \sinh^2 b  + e^{-4 j } } } .
    \label{eq:m}
\end{equation}

Results for symmetric degeneracies case $g_{+}= g_{-}$ and without phononic part repeat well-known behavior of conventional Ising model.

\subsection{Volume of the system and the correlation function}
\label{sec:Volume}

Let's calculate average effective volume of the molecular chain
\begin{equation}
    L = \sum_{n=1}^{N-1} \langle x_{n+1} - x_n \rangle = \sum_{n=1}^{N-1} \langle q_n \rangle.
\end{equation}
By the definition
\begin{equation}
    \langle q_n \rangle\! = \! \frac{1}{Z} \sum_{\langle s_1, \ldots, s_N \rangle}\! \iiint dq_1 \cdots dq_{N-1} q_n g_{s_1} \cdots g_{s_N} e^{-\beta E
    } 
\end{equation}
Integrating over the phonon degrees of freedom we get
\begin{equation}
    \langle q_n \rangle = \frac{\sum_{\langle s_1, \ldots, s_N \rangle} a_{s_ns_{n+1}} e^{-\beta H_{eff}  }}{\sum_{\langle s_1, \ldots, s_N \rangle}  e^{-\beta H_{eff} }}.
\end{equation}

Thus, the volume of the system $L = \sum_{n=1}^{N-1} \langle a_{s_ns_{n+1}} \rangle$.
We rewrite later expression as follows
\begin{equation}
     L = \sum_{n=1}^{N-1} \left( a_\epsilon + a_J \langle s_n s_{n+1}  \rangle + \frac{a_B}{2}  \langle s_n+s_{n+1}  \rangle  \right),
\end{equation}
where $a_\epsilon = \frac{1}{4} \left( a_{--} + a_{++} \right) + \frac{1}{2} a_{+-}$, $a_J = \frac{1}{4} \left( a_{--} + a_{++} \right) - \frac{1}{2} a_{+-}$ and $a_B = \frac{1}{2} \left( a_{++} - a_{--} \right)$.
Thus, the effective volume of system is connected with the average magnetization and the correlation function.
Average fictitious spin at the position $n$ corresponds to the local magnetization 
\begin{equation}
    \langle \hat{s}_n \rangle = \frac{ \Tr \hat{T}^{n-1} \hat{\sigma}_z \hat{T}^{N-n} \hat{R} }{ \Tr \hat{T}^{N-1} \hat{R} } .
\end{equation}
For calculating this expression we go to the eigenbasis of the transfer matrix.
All matrices except of $\hat{U}^{-1} \hat{\sigma}_z \hat{U}$ already have been calculated above.  
Therefore
\begin{equation}
    \hat{U}^{-1} \hat{\sigma}_z \hat{U} = 
        \begin{pmatrix}
        - m  
        & -1 - m \\
        -1 + m
        & m
    \end{pmatrix},
\end{equation}
where $m$ is the expression on the right side of the Eq.~(\ref{eq:m}).
We get local magnetization
\begin{equation}
    \langle \hat{s}_n \rangle = m + \frac{  C_{+-} e ^{-\frac{n-1}{\xi}}  + C_{-+} e ^{-\frac{N-n}{\xi}}  }{{ c_{+} + c_{-} e ^{-\frac{N-1}{\xi}}   } },
    \label{eq:m_exact}
\end{equation}
where coefficients are
\begin{equation}
    C_{-+} = C_{+-} = (m^2-1) (-\sinh w + e^{2j} \sinh b ) ,
\end{equation}
and the correlation length is $\xi = - \ln \frac{\lambda_{-}}{\lambda_+}$.
It is easy to see that since $\lambda_-<\lambda_+$, $\xi > 0$.
The average magnetization is
\begin{equation}
    \langle \hat{s} \rangle = m +  \frac{  C_{+-} + C_{-+}}{ N \left( 1 - e ^{-\frac{N}{\xi}} \right) \left(  c_{+} + c_{-} e ^{-\frac{N-1}{\xi}}  \right) }  .
    \label{eq:m_finite_N}
\end{equation}
In the thermodynamic limit average magnetization (\ref{eq:m_finite_N}) goes to magnetization given by the Eq.~(\ref{eq:m}).
We note that only average over all spins magnetization coincides with the classic Ising model result, while average of the individual spin distinct from the classic result due to the system boundary.
We see boundary effects do not vanish even in the thermodynamic limit.

The local correlation function $G_n(r)$ by definition is
\begin{equation}
    G_n(r) = \langle \hat{s}_n \hat{s}_{n+r} \rangle = \frac{ \Tr \hat{T}^{n-1} \hat{\sigma}_z \hat{T}^{r} \hat{\sigma}_z \hat{T}^{N-n-r} \hat{R} }{ \Tr  \hat{T}^{N-1} \hat{R} }.
\end{equation}
Therefore
\begin{multline}
    \langle \hat{s}_n \hat{s}_{n+r} \rangle = 
    m^2 + (1-m^2) \frac{ c_{+} e ^{-\frac{r}{\xi}} +  c_{-} e ^{-\frac{N-r-1}{\xi}} }{  c_{+} + c_{-} e ^{-\frac{N-1}{\xi}}  }
     \\ +  m C_{+-} \frac{  e ^{-\frac{n-1}{\xi}}-  e ^{-\frac{n-1+r}{\xi}} + e ^{-\frac{N-n-r}{\xi}} -  e ^{-\frac{N-n}{\xi}}  
     }{  c_{+} + c_{-} e ^{-\frac{N-1}{\xi}}  }.
    \label{eq:correlation_function}
\end{multline}
We get the correlation function in the thermodynamic limit
\begin{equation}
    G(r) = \frac{1}{N} \sum_{n=1}^{N-r-1} \langle \hat{s}_n \hat{s}_{n+r} \rangle = m^2 + (1-m^2) e^{-\frac{r}{\xi}}.
    \label{eq:correlation_function_thermodynamic}
\end{equation}
The average magnetization given by the Eq.~(\ref{eq:m_exact}) and the correlation function given by~(\ref{eq:correlation_function}) are exact. 
We see the average correlation function matches with the classic Ising model result~\cite{bellucci2013correlation} in the thermodynamic limit. 
Local correlation function (see Eq.~(\ref{eq:correlation_function})) has information about the edges of the system even in the thermodynamic limit.

Finally, we get expression
\begin{multline}
    L = (N-1) ( a_\epsilon +  a_B m +  a_{J} (m^2 + (1-m^2)  e ^{-\frac{1}{\xi}}  ))     
    \\ + a_B \frac{  C_{+-} + C_{-+}  }{{ c_{+} + c_{-} e ^{-\frac{N-1}{\xi}}   } } \left[\frac{1}{1 - e^{-\frac{N}{\xi}}} - \frac{1}{2}(1 + e^{-\frac{N-1}{\xi}}) \right] 
    \\ + a_{J} m \frac{C_{+-} ( 1 - e ^{-\frac{1}{\xi}} )}{ c_{+} + c_{-} e ^{-\frac{N-1}{\xi}}  } \left[ \frac{1}{1 - e^{-\frac{N-1}{\xi}}} + \frac{e^{-\frac{N-1}{\xi}}}{1 - e^{\frac{N-1}{\xi}}} \right].
    \label{eq:L_final}
\end{multline}
Expression (\ref{eq:L_final}) is exact and defines density $\rho^{-1} = L/N \to  a_\epsilon +  a_B m +  a_{J} (m^2 + (1-m^2)  e ^{-\frac{1}{\xi}}  ) $ in the thermodynamic limit.

\section{Spin-crossover chains}
\label{sec:Spin-crossover}

Here, we consider a general case where degeneracies of the pseudo-spin states are not equal $g_{+} \neq g_{-}$ and phonon corrections (\ref{eq:Ising_model_corrections}) are non-zero. 
This situation is typical for a number of molecular crystals, including spin-crossover materials.
Note that for spin-crossover molecular states, the relative degeneracy of the high-energy state in comparison with the low-energy one is large.

We introduce the equilibrium temperature $T_{eq}$ as a temperature when pseudo-spin states have equal occupations. 
This occurs when effective field vanishes $b = 0$.
Therefore we write
\begin{equation}
    T_{eq} = - \frac{ B}{k_B \left( \frac{1}{2} \ln g + \delta b \right) } .
\end{equation}
We note that for certain values of the external field $B$ and pseudo-spin degeneracies $g$, temperature $T_{eq}$ can be negative what means that for the given field and degeneracies there is no such temperature that pseudo-spin states would have equal occupations.

\begin{figure}[!htb]
    \centering
    \includegraphics[width=0.9\columnwidth]{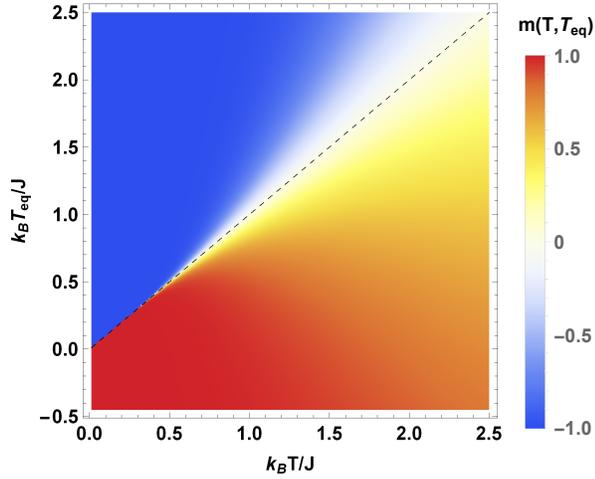}
    \caption{Average magnetization $m(T, T_{eq})$ for $\frac{1}{2}\ln g + \delta b = - 0.5$ and $\delta j = 0$. 
    Dashed line indicates condition $T=T_{eq}$. }
    \label{fig:Average_magnetization_phase_transition}
\end{figure}

\begin{figure*}[!htb]
    \centering
    \includegraphics[width=0.67\columnwidth]{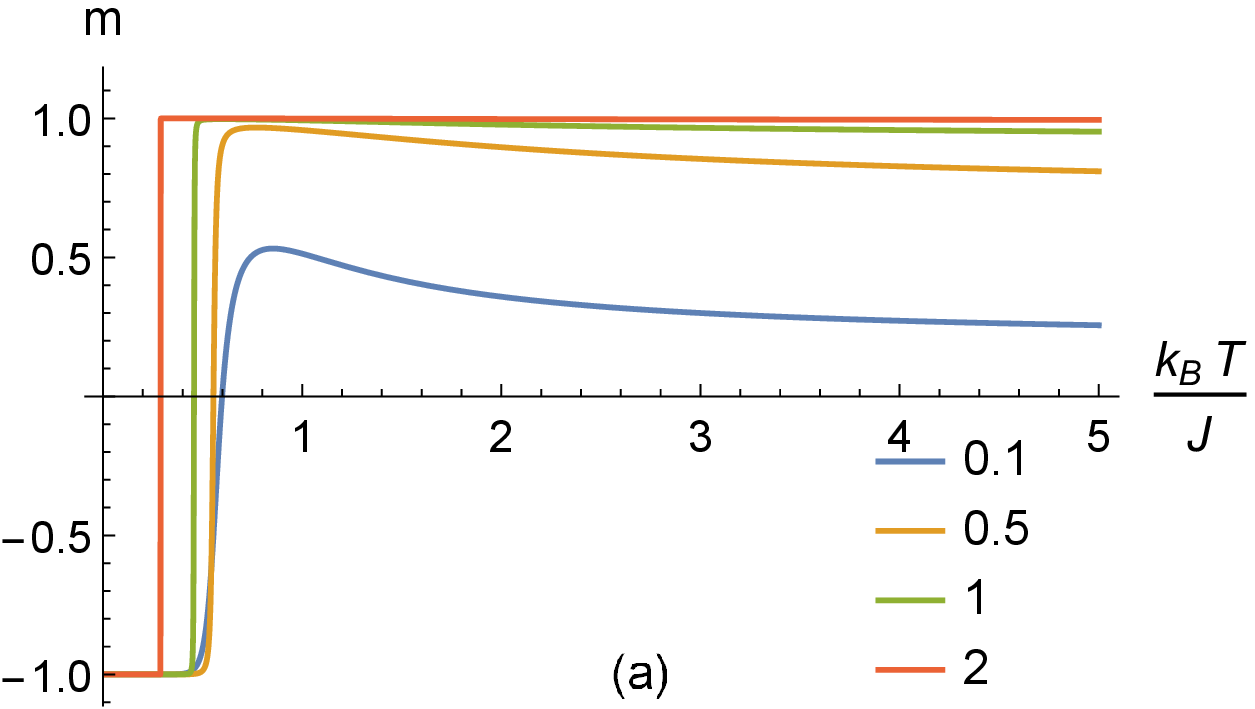}
    \includegraphics[width=0.67\columnwidth]{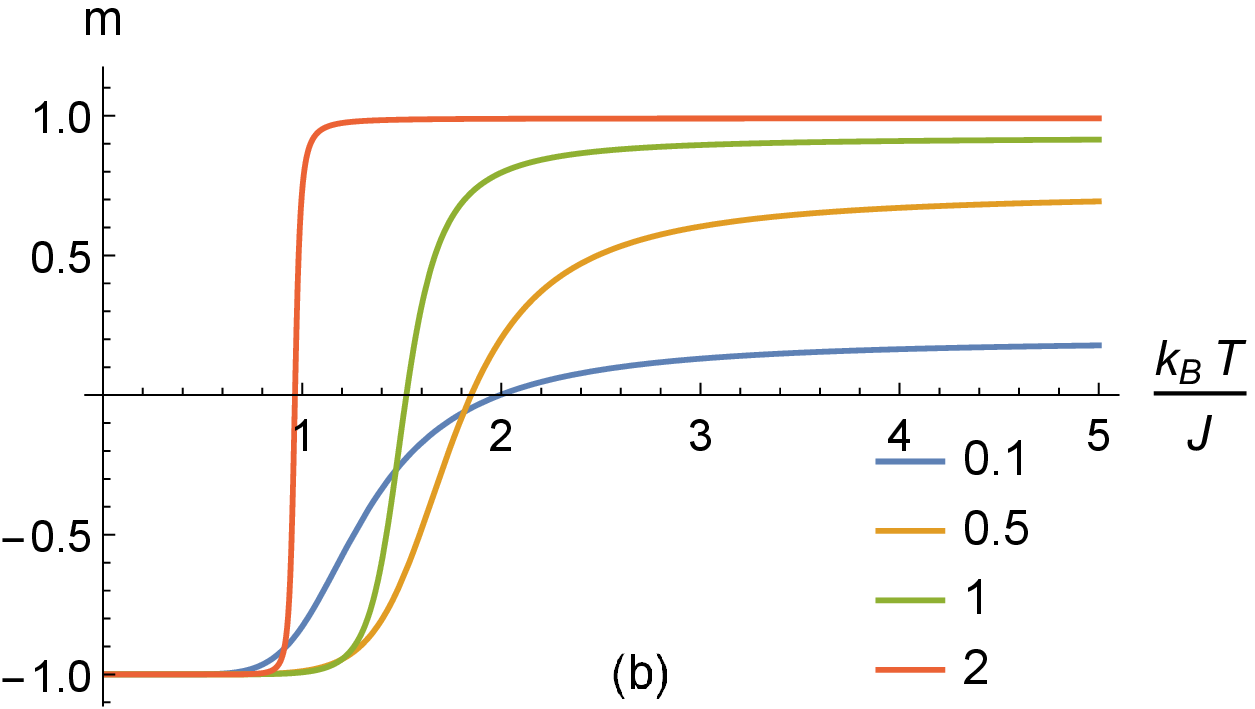}
    \includegraphics[width=0.67\columnwidth]{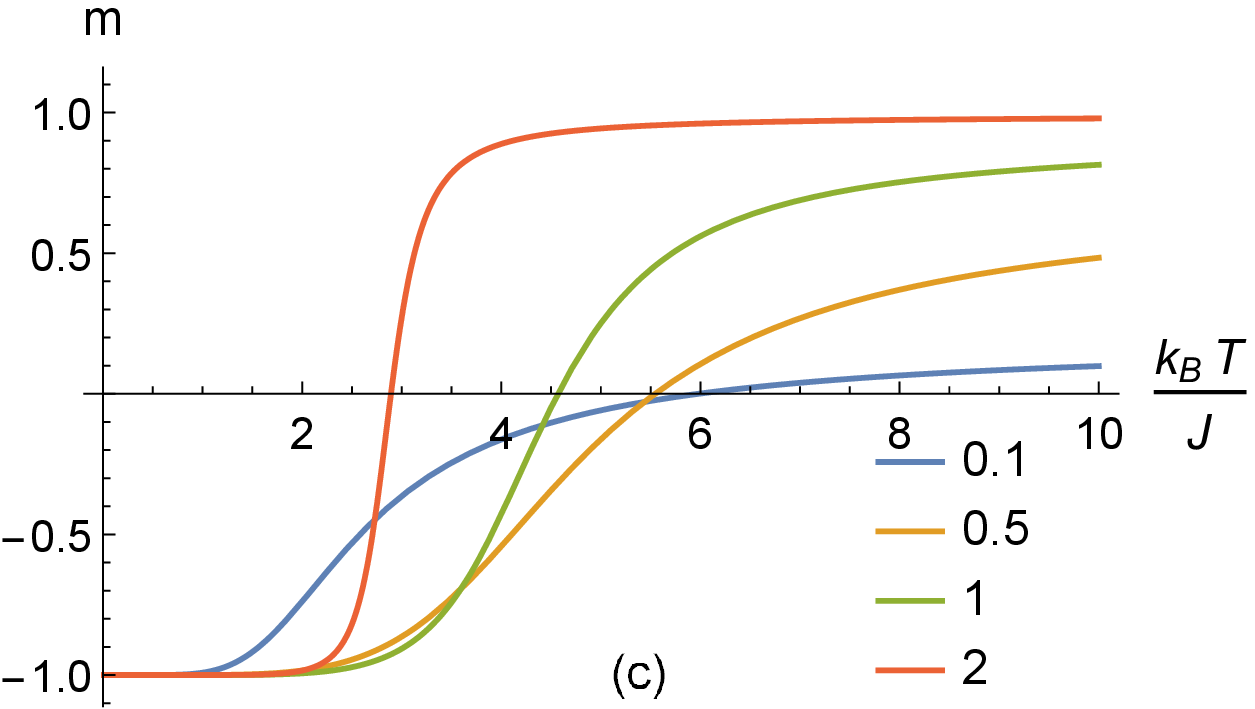}
    \caption{(a)-(c) Average magnetization as a function of temperature for $T_{eq}<T_{crossover}$, $T_{eq}=T_{crossover}$ and $T_{eq}>T_{crossover}$ and various values of the $\frac{1}{2}\ln g + \delta b = 0.1, 0.5, 1, 2$ and $\delta j = 0.35$.
    In the case $T_{eq}<T_{crossover}$ average magnetization has maximum at $T = T_0$.
    For $T_{eq}=T_{crossover}$, maximum is reached at $T_0 = \infty$.
    When $T_{eq}>T_{crossover}$ average magnetization is a monotonous function of temperature and has no extrema.}
    \label{fig:Average_magnetization_plots}
\end{figure*}
 
A spin-crossover material may abruptly change its macroscopic behavior, for example magnetization, when the external conditions, such as temperature, are varied.
The region where this happens is called abrupt crossover, and it marks a transition from one state to another. 
Dependence of average magnetization per spin as a function of temperature and equilibrium temperature $m\left(k_B T/J, k_B T_{eq}/J \right)$ for the fixed value $\frac{1}{2}\ln g + \delta b = 0.5$ is illustrated graphically in Fig.~\ref{fig:Average_magnetization_phase_transition}.
The dashed line corresponds to the condition $T = T_{eq}$, or equivalently $B_{eff} = 0$. 
In this case, the magnetization per spin $m = 0$.
The average magnetization is positive at any temperature when $T_{eq}$ is negative.
For a positive $T_{eq}$, $m>0$ when $T>T_{eq}$, and $m<0$ when $T<T_{eq}$, respectively.
We see that for a large $T_{eq}$, transition from $m=-1$ to $m=1$ is wide and smooth while at a small $T_{eq}$ this transition happens in the small region close to the point $T = T_{eq}$.
Such behaviour of the average magnetization in the limit $T_{eq} \to 0$ gives discontinuous behavior, and we have the phase transition at $T = 0$.
Therefore we talk about the crossover from smooth magnetization behaviour to jump-like.

We identify the crossover temperature in the following way.
At zero temperature the average magnetization is defined only by the sign of the effective magnetic field $m(T=0) = \sign(b)$, while in the high $T$ regime the average magnetization goes to value
\begin{equation}
    m(T \to \infty) = \frac{\sinh\left(  \frac{1}{2} \ln g + \delta b \right)}{ \sqrt{ \sinh^2 \left(  \frac{1}{2} \ln g + \delta b   \right) + e^{ - 4 \delta j} } } .
    \label{eq:m_high_T_limit}
\end{equation}
In the vicinity of $T = T_{eq}$, symmetry $\hat{s} \to - \hat{s}$ with $B_{eff} \to -B_{eff}$ exists and thus the average magnetization is an anti-symmetric function of $t = (T-T_{eq})/T_{eq}$.
Consequently for a small $T_{eq}$ the average magnetization changes abruptly from $m=-1$ to nearly $1$ and then goes to the high temperature asymptotic, while for large values of the $T_{eq}$ transition from $m=-1$ to the high $T$ asymptotic is smooth and has no maximum in-between.   
Therefore we identify the abrupt regime by existence of the local maximum of the average magnetization, and the temperature $T_0$ as the temperature at which the average magnetization has maximum.
The gradual regime we identify as a regime with no extrema in the average magnetization.
Thus condition for the existence of abrupt change in the average magnetization is 
\begin{equation}
    \left. \frac{\partial m}{\partial T} \right|_{T=T_0} = 0. 
    \label{eq:phase_transition_condition}
\end{equation}

Maximal equilibrium temperature $T_{eq}$ at which Eq.~(\ref{eq:phase_transition_condition}) has finite solutions for the $T_0$ is the crossover temperature.
The derivative is always positive and the magnetization $m$ is a monotonous function of temperature when $T_{eq}< T_{crossover}$.
For identifying the crossover temperature exactly we write
\begin{multline}
    \sign \left( \frac{\partial m}{\partial T} \right) =
    \sign \left( \frac{ k_B T_{eq} }{ 2 J } \left(  \frac{1}{2} \ln g + \delta b   \right) \right. \\ \left. -  \tanh \left( \left(  \frac{1}{2} \ln g + \delta b   \right) \left( 1 - \frac{T_{eq}}{T} \right) \right) \right).
\end{multline}
Abrupt crossover doesn't occur when $\frac{\partial m}{\partial T}$ has same sign at any temperature.
This might happen only when function $\tanh \left( \left(  \frac{1}{2} \ln g + \delta b   \right) \left( 1 - \frac{T_{eq}}{T} \right) \right)$ is smaller than $\frac{ T_{eq} \left(  \frac{1}{2} \ln g + \delta b   \right)}{ T_c (1 - \delta j)} $ at any finite temperature and equal to it at the infinite temperature.
Thus we get the crossover temperature 
\begin{equation}
    T_{crossover}  = \frac{2 J}{k_B} \frac{\tanh \left( \frac{1}{2} \ln g + \delta b \right) }{\frac{1}{2} \ln g + \delta b}.
    \label{eq:Tcrossover}
\end{equation}
We note that the function $\tanh(x)/x$ for all $x$ is smaller than one, and thus $T_{crossover} \leq \frac{2 J}{k_B}$. 
The maximal crossover temperature $T_{crossover} = \frac{2 J}{k_B}$ is reached when $\frac{1}{2} \ln g + \delta b \to 0$.
In Fig.~\ref{fig:Average_magnetization_plots}(a-c) dependence of the average magnetization on temperature for fixed $T_{eq}$ is plotted.
Three different cases are considered: (a) $T_{eq}<T_{crossover}$, (b) $T_{eq}=T_{crossover}$ and (c) $T_{eq}>T_{crossover}$.
\correct{
At high temperatures the magnetization goes to the asymptotic value given by the Eq.~(\ref{eq:m_high_T_limit}). 
At all panels (a-c) evolution of the magnetization is shown for four values of the $\frac{1}{2}\ln g + \delta b = 0.1, 0.5, 1, 2$ and $\delta j = 0.3$ being constant.
Corresponding curves of the same color from different panels share zero- and high temperature asymptotes.
The high temperature asymptotic value of magnetization $m(T \to \infty)$ goes to $0$ when $\frac{1}{2} \ln g + \delta b \to 0$ .
Asymptotic value $m=1$ is reached in the case of the large difference in the pseudo-state degeneracies ($g \gg 1$). }

Condition defining maximal magnetization is given by the Eq.~(\ref{eq:phase_transition_condition}). 
Therefore we get
\begin{equation}
    T_0 =  \frac{T_{eq}}{1 - \frac{1}{\frac{1}{2} \ln g + \delta b } \arctanh \left( \frac{ k_B T_{eq} }{ 2 J} \left(  \frac{1}{2} \ln g + \delta b   \right) \right)}  .
    \label{eq:T0}
\end{equation}
\begin{figure}[!htb]
    \centering
    \includegraphics[width=0.97\columnwidth]{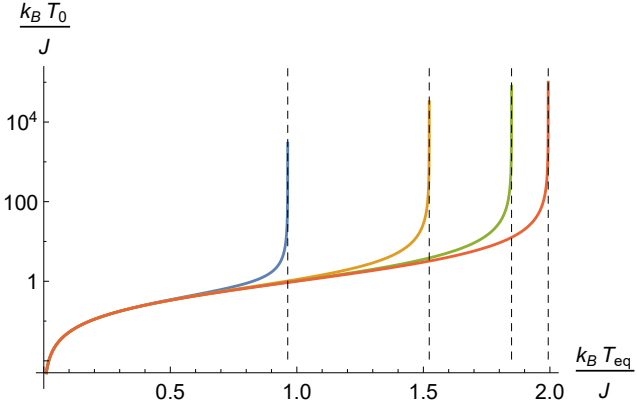}
    \caption{$T_{0}$ as a function of ratio $T_{eq}$ for various values of the expression $\frac{1}{2} \ln g + \delta b = 0.1, 0.5, 1$ and $2$ (right to left: red, green, orange, blue).
    Vertical dashed lines are asymptotes of the corresponding $T_{crossover}$ given by Eq.~(\ref{eq:Tcrossover}).}
    \label{fig:T0(B)}
\end{figure}
Dependence of $T_0$ from the $ T_{eq} $ for various values of $\frac{1}{2} \ln g + \delta b$ are plotted in Fig.~\ref{fig:T0(B)}.
In the limit $T_{eq} \to 0$, $T_0$ goes to zero as well.
With increase of the $T_{eq}$ goes to infinity as $T_{eq} \to T_{crossover}$.
In the region over the line defined by the Eq.~(\ref{eq:T0}) the derivative $\frac{\partial m}{\partial T}>0$ and in the region below $\frac{\partial m}{\partial T}<0$.
Eq.~(\ref{eq:T0}) can be rewritten explicitly in terms of the effective Hamiltonian parameters as following
\begin{equation}
    T_0  = - \frac{B}{k_B} \frac{1}{\frac{1}{2} \ln g + \delta b + \arctanh \frac{B}{J} }  .
\end{equation}
It is interesting to note that the average magnetization does depend on the phonon correction $\delta j$, while $T_0$ and $T_{crossover}$ do not depend on it.
\begin{figure}[!htb]
    \centering
    \includegraphics[width=0.95\columnwidth]{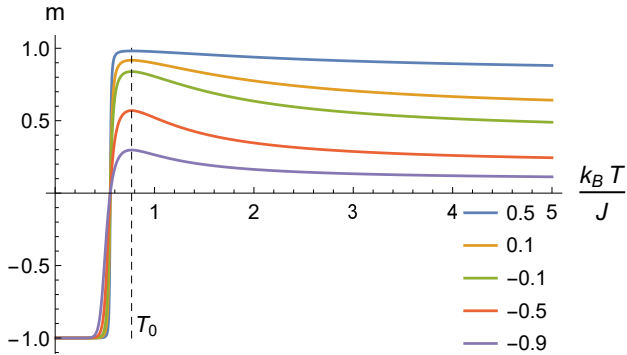}
    \caption{Average magnetization as a function of temperature for $T_{eq}=0.3 T_{crossvoer}$ and 
    various values of $\delta j = 0.5, 0.1, -0.1, -0.5, -0.9$.
    \correct{Maximum of magnetization is at $T = T_0$.}}
    \label{fig:Average_magnetization_dj}
\end{figure}
In Fig.~\ref{fig:Average_magnetization_dj} dependence of the average magnetization on temperature for fixed $T_{eq}=0.3 T_{crossvoer}$ is depicted.
The phonon interaction constant $\delta j$ was chosen from the set $0.5, 0.1, -0.1, -0.5, -0.9$.
\correct{With the increase of $\delta j$ we observe expected change of the high temperature asymptotes. 
This results in the difference of the magnetization change magnitude near the $T = T_{eq}$, but location of all characteristic temperatures remains the same.}

\correct{
The internal energy is $U = - \frac{\partial}{\partial \beta} \ln Z$
\begin{equation}
    U = E_0 + N \left( \frac{1}{2} k_B T - B \langle s \rangle - J G(1) \right),
\end{equation}
where the average local magnetization and the correlation function are given by the Eqs.~(\ref{eq:m_exact}) and (\ref{eq:correlation_function}).
In the thermodynamic limit these expressions reduce to the average magnetization $\langle s \rangle = m$ and the correlation function given by Eq.~(\ref{eq:correlation_function_thermodynamic}). 
\begin{figure*}[!htb]
    \centering
    \includegraphics[width=0.97\columnwidth]{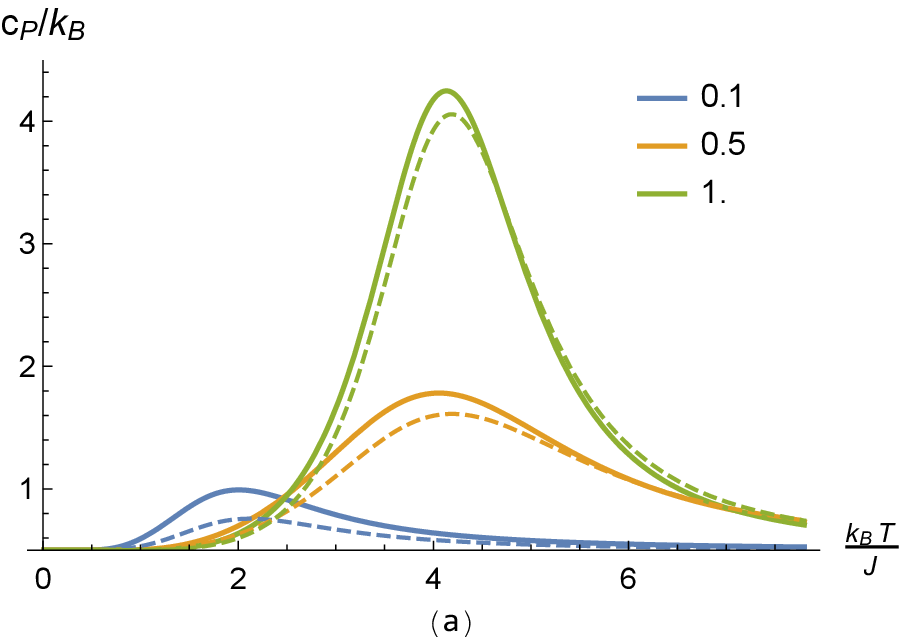}
    \includegraphics[width=0.97\columnwidth]{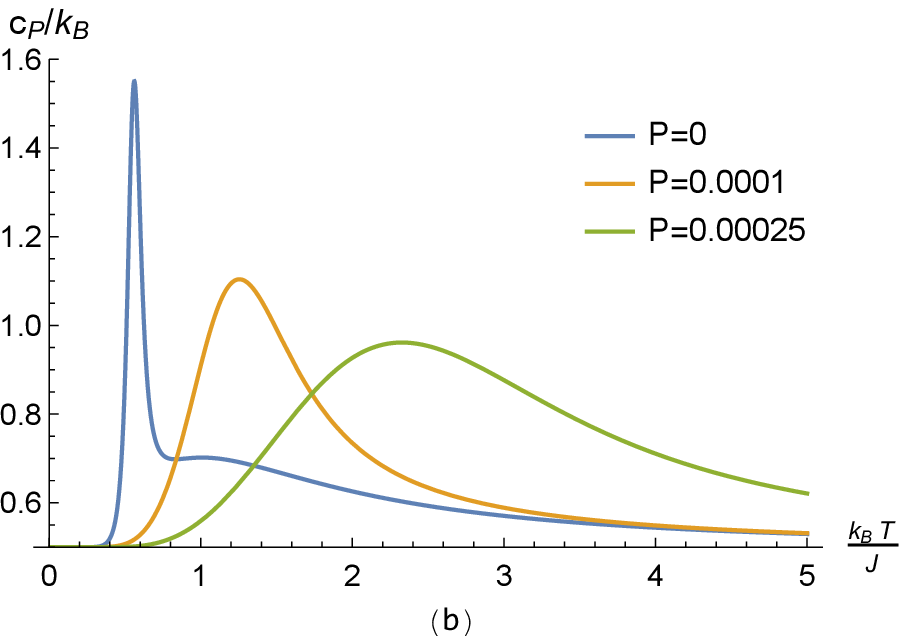}
    \caption{Specific heat capacity per particle $c_P$ (solid line) as a function of temperature and the $-B \frac{\partial m}{\partial T}$ derivative (dashed line). (a) $T_{eq} = 3 T_{crossover}$. 
    Parameters are chosen to be same as Fig.~\ref{fig:Average_magnetization_plots}(c).
    Maximum of the heat capacity is shifted from the maximum of the derivative $\frac{\partial m}{\partial T}$. 
    (b) $T_{eq} = 0.3 T_{crossover}$. 
    Parameters are chosen to be same as in Fig.~\ref{fig:Average_magnetization_plots}(a).
   }
    \label{fig:C_P}
\end{figure*}
Therefore the specific heat capacity per particle at constant pressure in the thermodynamic limit is 
\begin{equation}
    c_P = \frac{1}{2} k_B - B \frac{\partial m}{\partial T} - J \frac{\partial }{\partial T} \left(  m^2 + (1-m^2) e^{-\frac{1}{\xi}}  \right).
\end{equation}
In Fig.~\ref{fig:C_P} the specific heat capacity is plotted for two cases: (a) $T_{eq} = 3 T_{crossover}$ and (b) $T_{eq} = 0.3 T_{crossover}$.
These cases correspond to Figs.~\ref{fig:Average_magnetization_plots}(c) and (a).
Other parameters are fixed and chosen to be the same as for the orange curves in Fig.~\ref{fig:Average_magnetization_plots}.
At the left panel we observe the Schottky anomaly which transforms into large narrow peak at the right panel with increase of intermolecular interaction stiffness.
In the models with $J=0$ maximum of the heat capacity corresponds exactly to the maximum of function $\frac{\partial m}{\partial T}$~\cite{Boukheddaden2007}.
We note that in our model the location of the maximum of heat capacity is different from the maximum of the function $\frac{\partial m}{\partial T}$. 
This difference originates from the non-zero ferromagnetic interaction $J$ and is defined by the correlation function.
Another important effect we observe is the existence of the second maximum of the heat capacity when the crossover is abrupt.
In this case second peak has much smaller high and broader comparing to the narrow peak around $T_0$.
This effect is novel for 1D systems and was not present in similar 1D Ising-like models.}

\section{Summary and conclusions}
\label{sec:SUMMARY}

The recent progress of molecular magnetism towards low dimensions, with an increasing focus on applications, led to our research has been accented  on peculiarity of 1d spin crossover compound.
We have systematically studied the one-dimensional molecular chain with degenerate states and phononic interaction. 
Exact solutions by the method of transfer matrix modified for free boundary were obtained.
We have shown that there is an exact mapping between the elastic molecular chain  and Ising models of SCO materials. 
We have shown this explicitly for a one-dimensional model, but our approach can be extended to other elastic models and geometries.
We carefully consider the possible problems involving finite size effects and boundary conditions on the properties of the model which might be investigated in molecular chain.

An effective Ising-like Hamiltonian corresponding to the exact partition function of the system is obtained.
Moreover, this Hamiltonian and its structure have physical consequences, which implies that the effective Hamiltonian is much more than just a convenient mathematical construct.
The pivotal role in the transition in low-dimensional molecular structure can be physically attributed to the relative degeneracy of states.
The significant volume change in the cores molecules (SCO-molecules in the present case) leads to essential renormalization of the correlation function.
We have analyzed the regimes of the LS--HS crossover and identified if the crossover is abrupt or gradual for the specific parameters.
In the case of abrupt crossover we have showed possibility of existing two peak thermal dependence of the specific heat capacity. 

\section*{Data Availability}

The data that support the findings of this study are available from the corresponding author upon reasonable request.

\section*{References}
\bibliographystyle{unsrt}  
\bibliography{main}  

\end{document}